\documentclass[lettersize,journal]{IEEEtran}
\usepackage{amsmath,amsfonts}
\usepackage{algorithmic}
\usepackage{algorithm}
\usepackage{array}
\usepackage[caption=false,font=normalsize,labelfont=sf,textfont=sf]{subfig}
\usepackage{textcomp}
\usepackage{stfloats}
\usepackage{url}
\usepackage{verbatim}
\usepackage{graphicx}
\usepackage{cite}
\usepackage{multirow}
\usepackage{svg}
\usepackage{quoting}
\usepackage[dvipsnames]{xcolor}
\usepackage{hyperref}
\newcommand{\change}[1]{#1}
\newcommand{\remove}[1]{}

\begin{document}

\quotingsetup{vskip=3pt}

\newcommand{\confirm}[1]{\underline{#1}}

\newcommand{\extend}[1]{\underline{\underline{#1}}}

\title{“It just kind of shows that I went somewhere”:\\ An Exploratory Study of Fitness Data\\ \change{Visualization} Sharing}

\author{Mara Solen, 
    Thomas James Davidson, 
    Emily Wall,
    Tamara Munzner
    \IEEEcompsocitemizethanks{
        \IEEEcompsocthanksitem
        Mara Solen and Tamara Munzner are with the University of British Columbia, Department of Computer Science. E-mail: marasolen@gmail.com, tmm@cs.ubc.ca.
        \IEEEcompsocthanksitem
        Thomas James Davidson and Emily Wall are with the Emory University, Department of Computer Science. E-mail: thomas.james.davidson@emory.edu, emily.wall@emory.edu.}
    \thanks{Manuscript received XXX; revised XXX.}}

\markboth{TVCG Journal Submission, April~2026}%
{Solen \MakeLowercase{\textit{et al.}}: An Exploratory Study of Fitness Data Visualization Sharing}


\maketitle

\begin{abstract}
    The sharing of curated fitness data posts occurs frequently on fitness-focused social platforms such as Strava and on general social media platforms such as Instagram, which is a novel context for visualization. To better understand the process of sharing and designing fitness data posts, as well as the role of visualization within them, we \change{conduct} 18 semi-structured interviews with fitness data sharers. From our \change{thematic} analysis of the data, we find three novel characteristics \change{for visualization through our study of fitness data sharing:} (i) the role of visualization as \change{proof, (ii) the tension between individuality and conformity, and (iii) the complexity of design decisions with regards to effort and desire to share}. We also \change{discuss} a set of design implications, including a need for more options for visualizations for activities without routes, more user control in fitness data sharing platforms, and maintained ease of use while increasing customization options.
\end{abstract}

\begin{IEEEkeywords}
Visualization, fitness, social, communication.
\end{IEEEkeywords}

\section{Introduction}

\begin{figure*}[t]
    \centering
    \includegraphics[width=0.9\textwidth]{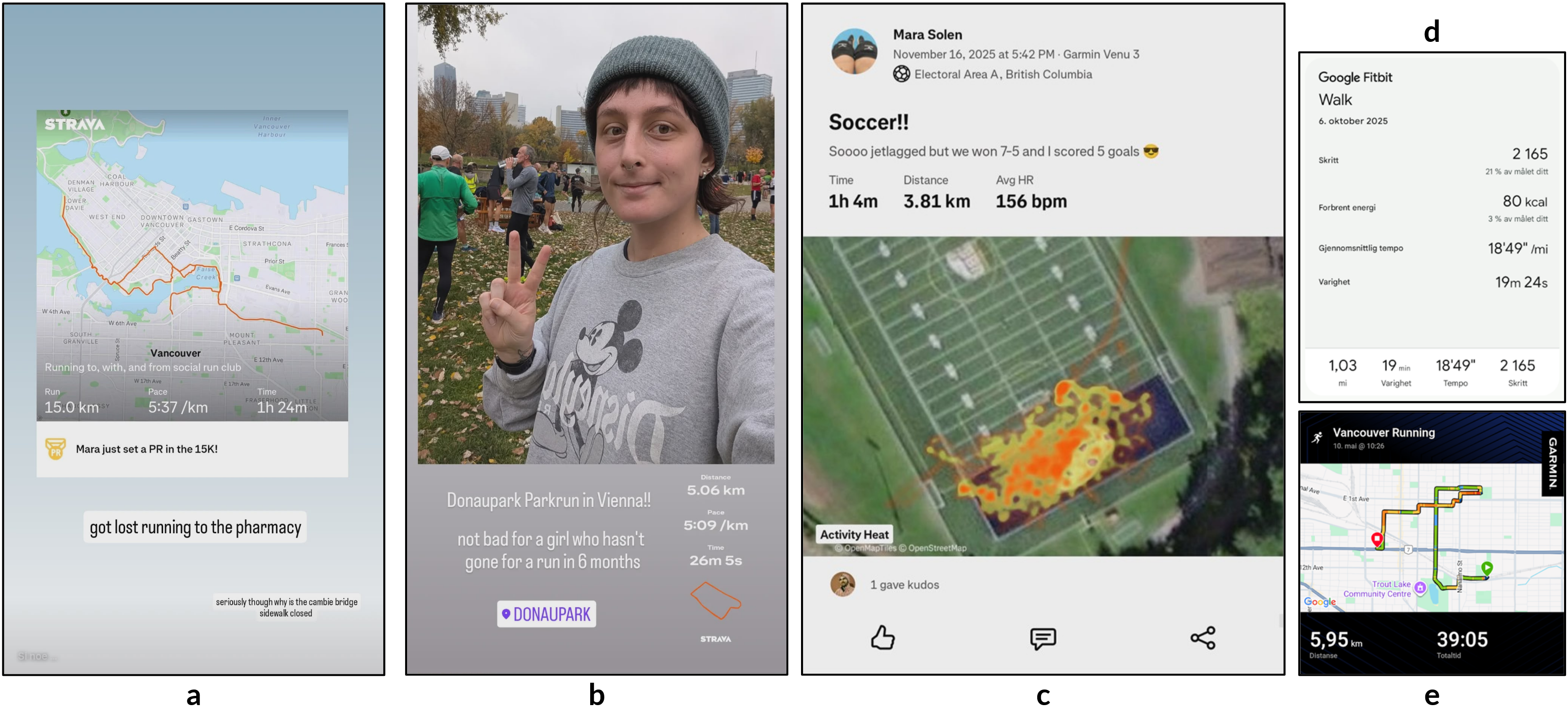} \\
    \caption{Five examples of FDS posts and exports from the first author. a) Instagram Story with a full map export from Strava. b) Instagram Story with a transparent-background outline-only route export from Strava. c) Strava in-app post with geographic heatmap. d) FitBit export showing 7 unique metrics. e) Garmin export showing route with embedded pace data}
    \label{fig:posts}
\end{figure*}

\IEEEPARstart{U}{sing} health data to support fitness progress is common \cite{bhargava2020opportunities} due to the ease of data collection through modern smart phones, watches, rings, and other devices. These devices record a variety of types of data, including the route an activity follows, the activity duration, the user's heart rate, and the user's calories burned, with different devices supporting different data for different activities. Applications to help wearers interpret their data, both from device developers themselves such as Garmin or third parties such as Bevel, are also readily available. These applications typically display the data in a variety of ways, including textually, numerically, and visually. They incorporate a variety of visualization idioms, such as geographic maps with \change{route lines}, line charts, and bar charts. Beyond self-analysis, many modern fitness data applications also support the sharing of fitness data \change{visualizations}, either through in-app, fitness-focused social ecosystems or through exports of graphics to be shared elsewhere on Instagram and other general social media platforms. Figure~\ref{fig:posts} shows five examples of fitness data posts and exported graphics.

The sharing of fitness data \change{visualizations} represents a unique context for visualization research. It is \textit{frequent}: users share very often, with some sharing multiple posts per day; it is \textit{curated}: sharers have the flexibility to customize their posts; and it is \textit{social}: the posts, including the visualizations, are shared on online social platforms. While there are some similar contexts, none share all three of these qualities. For instance, episodic overviews, like the music listening platform Spotify's ``Wrapped'' campaign and those inspired by it \cite{davidson2026spotify}, are often \textit{curated} and shared \textit{socially}, but not frequently, as many of these overviews are only available once a year. Similarly, performance in daily games like Wordle and games inspired by it are \textit{frequently} shared socially, but they are typically shared in a standard format without sharer curation. Likewise, the analysis of health data, such as sleep data \cite{liang2016sleepexplorer} or menstrual health data \cite{lin2024functional}, is often conducted \textit{frequently} and using \textit{curated} visualizations, although it is not shared socially.

While the \change{F}itness \change{D}ata \change{Visualization} \change{S}haring (FD\change{V}S) context is niche, we argue that it may be a harbinger of similar visualization contexts which will become more common in the future. \change{General Fitness Data Sharing (FDS)} \change{has become popular} due to the \change{early} availability of \change{GPS-enabled fitness watches and the popularity of discursive general and fitness-focused social media platforms, like Instagram and Strava \cite{rivers2020strava}}. Nowadays, improvements to devices and apps have led to them requiring significantly less technical proficiency, enabling a broader population to engage with them. We observe this and other similar trends occurring beyond FDS. In general, the availability and quantity of personal data is increasing, both through the number and diversity of features on wearable devices and the increase in passive collection of user interaction data on many platforms across multiple domains \cite{davidson2026spotify}. Moreover, the ability for people to access their data is also increasing with the advent of data protection laws which mandate access to users' personal data \cite{bowyer2022human}. 

We also note efforts to increase visualization ability in the general population, and in particular efforts to understand and improve skills such as the construction of visualizations \cite{ge2025avec, solen2022scoping}. Software for making expressive visualizations continues to become more readily available and easy to use. Finally, sharing data visualizations socially is becoming increasingly popular, as is visible through the explosion of platforms developing annual shareable episodic reviews such as Spotify's ``Wrapped'', Steam's ``Year in Review'' of video game play, YouTube's ``Recap'' of video watching, the social media platform LinkedIn's ``Year in Review'', and the social fitness platform Strava's ``Year in Sport'', which may facilitate reflection and sensemaking \cite{li2026platform}.

To better understand the design and design process of FD\change{V}S posts, and in particular the role of visualization, we conducted a semi-structured interview study \change{with 18 fitness data sharers, which we analyze using thematic analysis}.

Our primary contribution\remove{s} \change{is} the results of our \change{thematic analysis of the} semi-structured interview \change{transcripts, which are three novel characteristics of FDVS that may apply to similar visualization contexts}:
\begin{enumerate}
    \item \change{The role of visualization as evidence-based \textit{proof} of real-world activities, as opposed to communicating \textit{specific} insights about the data.}
    \item \change{The tension between expressing one's individuality and conforming to cultural norms and social pressure through design choices}.
    \item \change{The complexity of the design decisions that sharers make based on effort and desire to share}.
\end{enumerate}

Our secondary contribution is the suite of research artifacts we created, including a full codebook with both definitions and sample quotes for all codes, as well as anonymized full interview transcripts. These artifacts provide evidence for our analysis results and allow for alternative future analyses.

\section{Related Work}

Existing literature contains related work on a variety of connected topics including personal analytics, sociological perspectives of social FDS platforms, studies investigating the roles of individual post components such as specific metrics and images, and tools for designers of and sharers using FDS applications. However, no existing work investigates the intersection of these topics, which is the design and design processes, and in particular the role of visualization, of fitness data posts on popular existing platforms.

\subsection{Personal Informatics}

There is a large body of work exploring personal analytics for fitness \cite{fan2012spark, consolvo2008activity, li2012using} as well as other topics including music listening \cite{baur2010streams, dias2012interactive} and utility usage \cite{bartram2011smart, chetty2011my, costanza2012understanding}. This work typically focuses on users' personal interactions with their data, but can still offer some insights related to the \textit{personal data} aspects of personal data sharing.

Data agency, which refers to concerns about what information is collected and available to view, and effort, which refers to the amount of effort it takes to prepare visualizations, are two important concepts in personal informatics, and can influence whether tools are of interest to users \cite{huang2014personal}. Designing personal informatics tools that motivate users to sustain engagement is challenging, even when \remove{the tool is }intended to directly improve the users well-being by supporting their health \cite{moore2021exploring}. Disengagement with personal informatics\change{, either temporary or indefinite,} is called lapsing\remove{, which can be both temporary or indefinite,} and researchers have proposed many reasons for why it occurs \cite{epstein2015lived}. Temporary reasons include forgetting to track their data, facing challenges maintaining \remove{their} devices, intentionally choosing not to track specific instances, and intentionally suspending tracking behaviour due to a lack of interest or need. Some users indefinitely lapse for a variety of reasons, such as achieving their goals, a lack of benefit to tracking, or a loss of curiosity in the data.

While this research on personal analytics overlaps with our findings, it often focuses on data management and integration of the information into the person's life, and does not cover the sharing of personal data in depth.

\subsection{\change{FDS in} Sociology}

There is a significant body of work investigating social fitness platforms such as Strava from a sociological lens. Much of this work focuses on motivations for using these platforms, with some finding positive effects of Strava on an individual's relationship with sports \cite{evgenieva2024strava, franken2023kudos, russell2023if} and relationships with other people \cite{gui2017fitness}. This effect is amplified further when the individual receives positive social feedback \cite{franken2023kudos}. Existing work has also identified that Strava usage facilitates its own \change{online} community \cite{smith2014mobile}, within which different social classes form based on fitness level \cite{westlake2020if}, potentially leading to individuals feeling left out \cite{smith2014mobile}. Within these communities, there is social pressure to post about completed activities \cite{wozniak2017soil, smith2014mobile}, as omitting a post can lead to the fear of missing out \cite{smith2014mobile}. Felczak and Miros{\l}aw find that there are four frames of use for these platforms: social, hardcore, exploration, and training, and that different individuals have different motivations depending on their use \cite{felczak2025mobile, rivers2020strava}. This social aspect of these fitness platforms may describe why fitness data sharers are motivated to sustain their engagement, in contrast to users of other personal informatics tools as mentioned in the previous section.

This body of work also describes lower-level details of social fitness platform use. Users compare their posts with others within their networks \cite{evgenieva2024strava, williams2012king, rivers2020strava} to see what and how others are doing, \change{sometimes leading} to negative affect \cite{smith2014mobile}. Some users also use these platforms to facilitate self analysis \cite{williams2012king, couture2021reflections}, with that being their primary use. Self analysis can lead to a variety of affective outcomes \cite{smith2014mobile}, and in cases where the sharer is dissatisfied with their performance, they may remove the activity or make it private \cite{evgenieva2024strava, alqhatani2019there}. Sharers on these platforms are aware that others can see their posts, with some finding this awareness motivating \cite{rivers2020strava} while also being cognizant of how frequently they post \cite{wozniak2017soil}. 

While these sociological perspectives to social fitness platform use are  relevant and informative to our work and our findings align with many of them, they are confined to general platform use, and do not cover sharers' design decisions.

\subsection{Individual Post Components}

Existing work has investigated the impacts of individual post components, both on general social media platforms as well as social fitness platforms. Images, which are a common component across many social platforms, are found to significantly increase a viewer's motivation to engage with a post \cite{johnston2019motivating}, with photos getting the most likes and increasing the sense that the events described in the post really happened \cite{lowe2018thumbs}. Selfies have received significant attention from  \change{researchers}, who have found that they are used to seek attention, communicate information, archive information, and entertain others \cite{sung2016we}.

Metrics, such as heart rate, step count, or pace, are another post component investigated in existing work. Metrics can be used to provide simple, high-level summaries or complex, low-level details. Although fitness devices capture the information as low-level details, social fitness platforms typically restrict the sharing of metrics to high-level summaries. Researchers have investigated how to share low-level details socially, however while doing so they identified a key challenge of maintaining user privacy over their data \cite{epstein2013fine}.

While social fitness posts can include both media and metrics, these studies do not consider the design of the entire fitness post or how these components are juxtaposed or overlaid with others, nor do they discuss visualizations.

\subsection{Existing Tools \change{for Creating Shareable Visualizations}}

In 2020, Epstein et al. proposed Yarn, a tool for telling data-driven stories \cite{epstein2020yarn}. Their formative interview findings relate to our work, as users wanted to share a broad set of data types, such as images, text, and metrics, that emphasize aspects of the information that they found important. They also found that users preferred visual data formats and that sharers had a variety of goals with their posts, such as the goal of encouraging others. However, they found only modest success with the tool, and identified challenges, including that strict post templates constrained user creativity. 

Wang et al. investigated the use of personal-data-driven digital stickers on ephemeral social media posts \cite{wang2022snappi}. They discuss the differences between ephemeral and persistent posts, describing how ephemerality tends to be more private and playful, often documenting more mundane or everyday events. From their study, they find that the playfulness of the stickers made it easier and more enjoyable for users to share their information and start conversations with other users due to the perception of these stickers as less serious and more social than other platforms.
They also found that sharing personal data increased accountability in the sharers, that sharers customized their posts with stickers for specific audiences, and that sharers felt it was important to align with the stylistic norms of the sharing platform. While this study is informative, it does not investigate an established community focused on a specific topic, as the stickers covered a variety of topics and were shared to more general audiences.

Epstein et al.'s 2015 design framework for social sharing in personal informatics \cite{epstein2015nobody} also overlaps with our work. They identify many reasons why people share on social media, and provide a design framework for \remove{designing }tools for posting personal informatics based on an extensive review of prior literature. 

Our work differs from these existing tools in that we are investigating the use of existing, successful tools with established communities to understand low-level design decisions of users, with a particular emphasis on the visualizations\remove{ in the posts}.

\section{Methodology}
\label{sec:methodology}

\change{We conduct} a series of semi-structured interviews about the sharing practices and experiences of 18 participants who share their fitness data online, informed by a review of fitness data posts on Instagram Stories and of existing platform features. \change{We analyze the data using reflexive thematic analysis \cite{braun2019reflecting}}.

\subsection{Positionality and Bracketing}

The first two authors of this paper are embedded in fitness communities where FDS is common, including fitness class groups, sports teams, running groups, and online social fitness groups. Being embedded in these communities enabled the project through the identification of FDS as an interesting context for study through a visualization lens. It also enabled the sourcing of Instagram Stories about fitness data, and it provided access to and an increased trust level with interview participants. However, as both of these authors are regular sharers as well as visualization designers and researchers, they may think differently about the topic than those without visualization expertise, and as such have preconceptions and personal interests which are not reflected in the larger community \change{of fitness data sharers}.

We used a set of bracketing techniques to mitigate the deleterious effects of these preconceptions and personal interests \cite{tufford2012bracketing}. The first author, who was the primary data analyst in all analyses, conducted initial reflexive journaling, where she wrote out an extensive description of her ideas about FDS. During the data collection and analysis, \remove{both }the first and second authors wrote memos to document their developing understanding of the data. After each interview, \change{they} met to conduct peer debriefing, to further decant our developing understandings. Finally, all four authors met regularly to discuss the data collection and analysis, providing further opportunities to elicit and challenge preconceptions and biases.

\subsection{Preparation}

The first author reviewed public Instagram Stories with fitness data as well as features of FDS platforms to inform the design of the interview protocol and script.

In the fitness communities in which the first two authors are embedded, Instagram is a common place for the curating and sharing of fitness data information and  graphics sourced from FDS platforms. The first author has online connections with local fitness leaders such as fitness class instructors and run club leaders who publicly share their own fitness data posts and re-share those of others in their fitness communities, with or without additional curation. The first author opportunistically took screenshots of Instagram Stories from these leaders from April 2024 to September 2025, leading to a collection of 54 screenshots of Stories created by 37 unique users. She then reviewed the screenshots to gain a preliminary understanding of what people include in fitness data posts. \change{In Canada, where the first and last authors are located, research using public artifacts is permitted by the national Research Ethics Board. However, since Instagram Stories are ephemeral and not intended to exist in perpetuity, we do not share the screenshots to respect the original posters \cite{fiesler2018participant}.}

The first author then explored the sharing features of four fitness data sharing platforms which were most commonly used in the set of Instagram Stories: Strava, Garmin Connect, Apple Health, and FitBit. She collected screenshots of the shareable outputs and wrote descriptions of the features to further inform our interview design. See supplemental materials for a description of the platform features.

\subsection{Data Collection}

\change{We developed our semi-structured interview script and protocol based} \remove{Based }on author expertise and the reviews of the Instagram Stories and of the platform features\remove{, we developed our semi-structured interview script and protocol}. The interview script had three sections: 
\begin{enumerate}
    \item participant demographics and orienting questions about fitness level, fitness activities, and general social media use\change{, as well as questions about what specific fitness communities, such as run clubs or online fitness groups, the participants were involved in};
    \item questions about participant sharing practices and specific examples of shared posts;
    \item and prompts about different features and points of dissatisfaction noted by previous participants.
\end{enumerate}
We made small modifications to the interview script throughout the interview phase to improve flow and clarity for future participants. We asked for screensharing when we were unfamiliar with what the participant was describing. 

We recruited interview participants through social media. The first author posted the flyer on Instagram and asked local fitness leaders to share it to their connections. We also advertised the flyer on the first and fourth authors' university department's Slack. Our focus on the \change{first author's} local community \change{of Vancouver-based individuals who post their fitness data} allowed us to reach faster saturation, and is intended to describe a context grounded in a particular community rather than being globally representative.

Individuals who saw the flyer and were interested in participating in an interview filled out a pre-screening survey to verify that they met inclusion criteria. To be included in the study, respondents had to be nineteen years old or older, speak English, and have shared fitness data socially before. We also asked users to briefly characterize their fitness data sharing, such as what tools they used, how often they posted, and what fitness activities they partook in. We chose our initial participants to achieve diversity in these characterizations, and then later invited additional participants to further saturate areas where we found we lacked data based on memos and internal discussions. We excluded participants with established personal relationships with the interviewers. We conducted 2 pilot interviews with colleagues to refine our interview protocol and script which we exclude from the analysis. We then conducted 18 more interviews before reaching saturation. \change{One of our interview participants had shared Instagram Stories that we used in preparation for the interviews}. Participants included 9 women, 6 non-binary people, and 3 men ranging from 22 to 53 years of age who were all \change{currently or previously }based in \remove{or recently moved from} Vancouver, Canada. \change{Due to the exploratory nature of this study, we did not find it necessary to achieve a more gender-balanced set of participants. Further, men are historically over-represented in sports and exercise research \cite{cowley2021invisible}, and we see the large number of female and non-binary participants as a strength of our study}. Their \change{26} different fitness activities included running (15), gym (10), walking (8), cycling (5), and 24 others. \change{The bias towards running, walking, and cycling in our sports also reflects the population of our study, as most fitness apps support sports with location data significantly more than other sports.} Table~\ref{tab:participant-demographics} provides participant context to support our results, including participant age, gender, sports played, devices used for tracking fitness, and platforms used to share about fitness activities

\change{
\begin{table*}[]
    \centering
    \raggedright
    \begin{tabular}{p{0.02\linewidth}|p{0.025\linewidth}|p{0.044\linewidth}|p{0.41\linewidth}|p{0.20\linewidth}|p{0.165\linewidth}}
        \textbf{P\#\#} & \textbf{Age} & \textbf{Gender} & \textbf{Sports} & \textbf{Devices} & \textbf{Platforms}  \\ \hline
        P01 & 31 & W & running, hiking, pilates, yoga, cycling, spin, walking & Garmin watch & Strava, Instagram \\
        P02 & 29 & M & running, boxing, walking & Apple watch & Strava, Instagram \\
        P03 & 27 & W & running, hiking, basketball, gym, skiing, yoga, pilates & Garmin watch, Casio watch & Strava \\
        P04 & 30 & W & running, cycling, swimming, climbing, gymnastics, walking & Garmin watch, FORM \remove{smart swim} goggles & Strava, Instagram, Garmin \remove{Connect} \\
        P05 & 24 & W & running, roller skating & smartphone & Strava \\
        P06 & 24 & NB & running, gym & Galaxy watch, smartphone & Strava \\
        P07 & 29 & W & running, gym, boxing, circuit training & smartphone, Oura ring & Strava, Instagram \\
        P08 & 22 & W & volleyball, gym, hiking, walking & smartphone, Apple watch & Strava, Instagram \\
        P09 & 37 & NB & running, climbing, cycling, tennis, kayaking, walking & Chorus watch & Strava \\
        P10 & 27 & M & running, skiing & smartphone & Strava \\
        P11 & 31 & NB & cycling, climbing, gym, walking & Garmin watch & Strava \\
        P12 & 26 & W & running, swimming, cycling, walking & Garmin watch & Strava, Instagram \\
        P13 & 25 & NB & running, gym, roller skating, skateboarding, ice skating, hockey & smartphone & Strava \\
        P14 & 31 & W & running, swimming, gym & Garmin watch & Strava, Instagram \\
        P15 & 36 & M & running, pickleball, golf, gym, spin & Garmin watch & Strava, Garmin \remove{Connect} \\
        P16 & 53 & W & boxing, gym, walking & Apple watch & Strava, Instagram \\
        P17 & 34 & NB & running, circuit training & Garmin watch & Strava, Gaia \\
        P18 & 29 & NB & hockey, basketball, dodgeball, running, pickleball, softball, gym & Garmin watch & Strava \\
    \end{tabular}
    \caption{\change{Demographics of the interview study participants. W = woman, M = man, NB = non-binary.}}
    \label{tab:participant-demographics}
\end{table*}}

The first two authors participated in all interviews using pair interviewing, where one interviewer acts as the ``driver'' and focuses on asking the questions on the script, while the other interviewer acts as the ``navigator'' and focuses on taking notes and asking targeted follow-up questions \cite{akbaba2023two}. We conducted the interviews remotely over Zoom. The interviews lasted on average 60 minutes. We provided a \$25 CAD gift card to a fitness-related store for participating in the study. Zoom provided automatic transcriptions of the interviews which we manually edited where necessary. Our final interview dataset includes around 193,000 total words and over 1,000 minutes of interview audio. We supplement this dataset with memos taken by the interviewers both during each interview and in the debrief discussion immediately following each interview.

See supplemental materials for the full interview protocol and script as well as de-identified interview transcripts.

\subsection{Data Analysis}


\change{As per reflexive thematic analysis \cite{braun2019reflecting}, we}\remove{We} conducted a multi-stage coding process to analyze the Interviews. The first author \change{familiarized herself with} all data, including memos and transcripts from the interviews. She then conducted in vivo coding \cite{manning2017vivo} of the interview transcripts with the NVivo tool, highlighting snippets for further analysis. She created an initial codebook based on her initial review of the data.

She then coded the interview transcripts in depth, producing a \change{hierarchical codeset with three levels of codes}. \remove{The properties are a middle level in between the category level and the code level, and serve to characterize the category. One code can be included in many properties and one property can involve many codes. }The second author, who participated in all interviews, then joined the refinement iterations on the codebook, discussing with the first author and providing feedback which led to refinements \cite{elmqvist2012patterns}. Finally, the third and fourth authors joined the final refinement iterations of the codebook. As these two authors did not participate in the interviews, their feedback primarily led to clarity refinements.

To determine saturation, we recorded the number of new \change{mid-level codes} that emerged with each new interview. We chose \change{mid-level codes} over \change{low-level }codes as the granularity of codes meant that several new \change{low-level }codes could emerge without informing new \change{mid- or high-level codes}.
As we found only one new \change{mid-level code} from the last five interviews, we determined we had reached saturation\change{, aligning with previous work describing 12 as a common number of interviews required to reach saturation \cite{guest2006many}}. The final codebook contains 11 \textbf{\change{high-level codes}}, spanning 58 \textbf{\change{mid-level codes}}, and 130 total \textbf{\change{low-level codes}}. Each code is characterized by a definition, sample quote, and description of which \change{participants' interviews} the code applies to. Each \change{mid-level code} is characterized by a definition, list of associated \change{low-level codes}, and which participant the \change{mid-level code} was first noticed in. \remove{The categories are further aggregated into 4 higher-level \textbf{category groups}: purposes, posts, perceptions, and desires. }The entire codebook, from the first iteration to the \change{fifth} and final iteration, is included in supplemental materials. 

\change{Informed by the codebook, we then constructed a set of three themes that describe the novel characteristics of the FDVS context.}
\remove{We conducted a reflective second stage of analysis, drawing from the entire codebook, to produce a set of three novel characteristics of FDS and a set of three design implications for FDS platforms.}

\begin{table*}[]
    \centering
    \raggedright
    \begin{tabular}{p{0.08\linewidth}|p{0.03\linewidth}|p{0.20\linewidth}|>{\noindent\raggedright}p{0.10\linewidth}|p{0.49\linewidth}}
        \hline
        \textbf{High-Level Code} & \textbf{Label} & \textbf{Mid-Level Code} & \change{\textbf{Previous Work}} & \textbf{Definition} \\ \hline
        \multirow{7}{13em}{\textbf{Purposes} \newline for Sharing}              & A1 & Shares for personal benefit                      
        & \cite{franken2023kudos, smith2014mobile, rivers2020strava, russell2023if, couture2021reflections} 
        & They share because it benefits them, the sharer. \\
                                              & A2 & Shares for others                                & \cite{rivers2020strava, wozniak2017soil} & They share because it benefits the people who see the shared post. \\
                                              & A3 & Shares for sense of community                    & \noindent\cite{williams2012king,smith2014mobile,rivers2020strava} & They share to build, maintain, and be reminded of their community. \\
                                              & A4 & Shares for access                     & \cite{couture2021reflections} & They share to have access to the data and provide access to others. \\ \hline
        \multirow{7}{13em}{\textbf{Posts} \newline Goals}        & B1 & Cares about quality of posts                     & \cite{wang2024exploring} & They care about the perceived quality of their posts. \\
                                              & B2 & Cares about quantity of posts                    & \cite{wozniak2017soil} & They are hesitant to share too many posts, or too many similar posts. \\
                                              & B3 & Cares about feed contribution & \cite{wozniak2017soil} & They care about how the posts they make appear on and contribute to others' feeds. \\
                                              & B4 & Seeks summaries                                  & \cite{epstein2013fine} & They want high-level summaries, emphasizing they did the activity, not fine-grained details. \\
                                              & B5 & Wants to represent their values                  & \cite{seidman2013self} & They want their posts to represent their values. \\ \hline
        \multirow{5}{13em}{\textbf{Posts} \newline Effort}       & C1 & By platforms/types of posts                      & \cite{wang2022snappi} & The platform and type of post (temporary, permanent) impact effort required to make post. \\
                                              & C2 & By activity                                      &  & Certain activities take more effort to post about than others. \\
                                              & C3 & Comparing to others                              &  & They compare the amount of effort they put in to others they know. \\
                                              & C4 & Not posting due to effort                        &  & They do not make as many posts as they'd like due to the effort required. \\ \hline
        \multirow{6}{13em}{\textbf{Posts} \newline Frequency}            & D1 & General practice                                 & \cite{evgenieva2024strava, alqhatani2019there} & They follow a typical pattern in how often and what activities they post. \\
                                              & D2 & Avoiding sharing                                 & \cite{evgenieva2024strava, alqhatani2019there} & They avoid sharing in general or just specific activities. \\
                                              & D3 & By performance                                   & \cite{evgenieva2024strava, alqhatani2019there} & Their performance impacts their choice of whether or not to post. \\
                                              & D4 & By activity                                      &  & They limit their postings to certain types and/or intensities of activities. \\
                                              & D5 & By platform                                      &  & Their posting patterns are different on different platforms. \\
                                              & D6 & Changes to frequency                             &  & Their posting patterns changed over time. \\ \hline
        \multirow{8}{13em}{\textbf{Posts} \newline Format}       & E1 & Sharing qualitative data                         & \cite{wozniak2017soil} & They share qualitative data like textual descriptions. \\
                                              & E2 & Sharing quantitative data                        & \cite{epstein2013fine, wozniak2017soil} & They share quantitative data like performance metrics. \\
                                              & E3 & Sharing visuals                                  & \cite{johnston2019motivating, lowe2018thumbs} & They share visuals including visualizations and media. \\
                                              & E4 & About defaults                                   &  & Participants describe whether 
                                              they change default information in their posts. \\
                                              & E5 & Platform comparisons                             &  & They use different posting formats on different platforms. \\
                                              & E6 & By performance                                   &  & Their post format is dependent on their performance. \\
                                              & E7 & Format changes                                   &  & The format of their posts has changed over time. \\ \hline
        \multirow{5}{13em}{\textbf{Posts} \newline Content}      & F1 & Sharing context                                  &  & They share details about their activity that aren't captured automatically. \\
                                              & F2 & Relating to performance                          &  & They share information about and reasons for their performance. \\
                                              & F3 & Sharing content for others                       &  & They include content that is meant to benefit others. \\
                                              & F4 & Sharing non-activity content                     &  & They share content that is not directly related to the activity. \\
                                              & F5 & Changes to content choices                       &  & Their content choices changed over time. \\ \hline
        \multirow{6}{13em}[-0.5em]{\textbf{Perceptions} \newline of Viewers}  & G1 & Caring about sharing                & \cite{williams2012king} & Participants describe if they think people, them and others, care or not about fitness posts. \\
                                              & G2 & About specific people                            &  & Perceptions about how they think specific people will respond. \\
                                              & G3 & Fears                                            & \cite{williams2012king} & They have fears about posting. \\
                                              & G4 & Potential reactions                              &  & They consider potential reactions from others when posting. \\
                                              & G5 & Engagement                                       &  & They consider how engaged viewers will be by their posts. \\
                                              & G6 & Changes to feelings                              &  & Their feelings about fitness data sharing have changed over time. \\ \hline
        \multirow{6}{13em}{\textbf{Perceptions} \newline of Metrics \newline}        & H1 & Metric relevance                                 &  & They find certain metrics more relevant to them than others. \\
                                              & H2 & Liking metrics                                   &  & They like certain metrics. \\
                                              & H3 & Disliking metrics                                &  & They dislike certain metrics. \\
                                              & H4 & Uses of metrics                                  &  & They use metrics to describe certain aspects of their activities. \\
                                              & H5 & Preferences for metrics                          & \cite{epstein2013fine} & They have preferences for what metrics to share. \\ \hline
        \multirow{6}{13em}{\textbf{Perceptions} \newline of \newline Visualizations \newline} & I1 & Describes activity                               & \cite{tong2018storytelling} & They find that visualizations help to describe what the activity was. \\
                                              & I2 & General enjoyment                                & \cite{lowe2018thumbs, johnston2019motivating} & They enjoy visualizations. \\
                                              & I3 & Information density                              & \cite{munzner2014visualization} & They find certain visualizations to have high or low information density. \\
                                              & I4 & Using maps to find specific info           & \cite{chevalier2013using} & They find  maps are useful for sharing specific information like distance, routes, \& location. \\ \hline
        \multirow{6}{13em}{\textbf{Perceptions} \newline of Platforms}& J1 & Thoughts on Strava                               & \cite{smith2014mobile} & Participants share thoughts on Strava in general. \\
                                              & J2 & Garmin is challenging to use                     &  & They find that Garmin is hard to use. \\
                                              & J3 & Compares Garmin and Strava                      &  & They make comparisons between Garmin and Strava. \\
                                              & J4 & Compares Strava and \mbox{Instagram}                   &  & They make comparisons between Strava and Instagram. \\
                                              & J5 & Aesthetics                                       &  & They discuss the general and sharing-specific aesthetics of platforms. \\
                                              & J6 & Purpose                                          &  & They discuss the purpose they associate with different platforms. \\ \hline
        \multirow{9}{13em}{\textbf{Desires} \newline for Features \newline}& K1 & Improved customizability    &  & They want more customizability in existing features of fitness data sharing tools. \\
                                              & K2 & Flexibility with metrics                         &  & They want more flexibility with what metrics they can share. \\
                                              & K3 & Diversity to other sports                        &  & They want more diverse posting options for sports aside from running/cycling. \\
                                              & K4 & Diversity for sharing visuals                    &  & They want more diverse visualization options for sharing. \\
                                              & K5 & More features for Strava sharing           &  & They want new sharing features in Strava. \\
                                              & K6 & Doesn't know how to improve              &  & They don't know how to improve fitness data sharing platforms. \\ \hline
                                              \multicolumn{4}{c}{} \\
    \end{tabular}
    \caption{The full table of high- and mid-level codes. \change{Previous Work = previous work related to the mid-level codes.}}
    \label{tab:properties}
\end{table*}

\remove{\section{Results}}
\label{sec:codebook}

\section{\change{Results: Novel Characteristics}}

Many aspects of the FD\change{V}S context align with general communicative visualization contexts\change{, such as} the importance of engagement and understandability \remove{of the posts reflects goals in contexts such as informal education }\cite{solen2024delve, sanchez2023effect}. However, we identified three novel characteristics \change{for visualization through our study} of FDS: the role of visualization as proof, \change{the tension between individuality and conformity, and the complexity of design decisions based on effort and desire to share}.

\subsection{Visualization \change{Existence} as Proof of Activity}

\remove{For FDS posts with visualizations, we found that the common visualization tasks \cite{brehmer2013multi} may not be relevant, as multiple participants described their use of visualizations without mentioning any particular insight from the data. However, visualizations are still necessary in fitness data posts, as our interview participants found posts without visualizations to be less appealing, and they were less likely to post about activities without visuals. Beyond appeal, the visualizations provide more evidence that the activity was real than simply describing the activity in text or with numbers. For activities with routes, the visualizations generally show that the individual actually went outdoors for the activity rather than staying indoors and using a stationary machine like a treadmill or stationary bike. The data in the visualization, collected via a third party in the form of the fitness device, acts as a stamp of confirmation.}

\change{Our participants described multiple reasons for why visualizations improved their fitness posts. Some described them as simply being aesthetically pleasing, while others used them for deeper purposes like telling a story about their experience during the activity. One described how sharing maps enabled a user to show others ``\textit{the journey that [the user] took}'' (P08), using the example of a remote hike with many switchbacks. Participants also said that they liked the geographic maps with route outlines that Strava and other apps generate because it is useful for sharing information such as the distance, route, and location of the activity. One participant described how maps were particularly useful for helping viewers to truly comprehend the distances of long runs, saying that
\begin{quoting}
    ``\textit{It's nice, like, to get an idea of how long something really is. Because, like, when you think of, like, 50K, you're like, oh, that's not that long, but then if you think of, like, point A to point B, then you're like, oh my god, that's actually so long}'' (P04).
\end{quoting}
Accurately communicating metrics such as distance is important because they are helpful for providing ``\textit{a good benchmark for [...] how you're doing}'' (P04). P15 described how he had planned a vacation without telling some friends, and, from their perspective, he had simply disappeared while he was away. However, including the map in his fitness posts revealed where he had been to those friends.}

\change{These fitness-app-generated visualizations are so valuable that participants make decisions about which activities to post based on their availability. P04 clearly stated this idea:
\begin{quoting}
    ``\textit{Any run, like, any cycle, anything with a map, I'll always post}''.
\end{quoting}
Because of this value, participants described exporting visualizations to post on other platforms, either by using built-in export tools or by simply taking a screenshot of the relevant visualizations. As implied by P04, only activities with location data have visualizations that appear in default Strava posts. Because of this gap and the importance of visualizations, participants sometimes use other apps to generate visualizations, with P04 describing how their smart goggles can generate a visualization of the different strokes they did during a swim and automatically upload it as an image to Strava.}

\change{Geographic maps are not the only visualizations that Strava and most other fitness apps can generate in an easily-shareable way. Strava also supports users in exporting isolated route outlines: lines showing the path of the route without any geographic details in the background, as shown in the bottom right corner of Figure~\ref{fig:posts}b. While these route outlines can still portray some aspects of the activity experience, such as switchbacks, they provide significantly less information due to their lack of geographic context. For example, it is virtually impossible to gauge the total distance of a route using a route outline in isolation. As P09 stated: ``\textit{[a route outline] doesn't really tell you anything by itself.}''}

\change{Despite this lack of information, our participants still found route outlines valuable beyond simply looking ``\textit{cool}'' (P06). Route outlines support sharers primarily by providing visual proof that they actually did an activity:
\begin{quoting}
    ``\textit{It's like a stamp of, like, I did this [...] And so, I do feel like it's more real with that little map, or the little outline without the map.}'' (P05)
\end{quoting}
These location-based visualizations do not only provide proof that a person did an activity; they can also provide proof that the person went somewhere interesting, rather than ``just'' working out indoors. P02 described how this form of proof impacted his posts about running:
\begin{quoting}
    ``\textit{It just kind of shows that, like, I went somewhere, you know what I mean? Like, as opposed to just, like, whatever I was running on a treadmill.}''
\end{quoting}
Some participants even described a preference for these less-informative route outlines. We primarily observed this sentiment in participants who preferred sharing \change{only }high-level summaries of their data\remove{ without the low-level details}, often for privacy reasons. Geographic maps are revealing in terms of both location, which could be tied to the sharer's home or workplace, as well as distance, which could be related to the sharer's fitness level. Meanwhile, route outlines do not expose this data.}

\change{While visualizations about fitness activities are naturally valuable for communicating insights about the data through standard visualization tasks, their ability to act as proof that something happened is novel in visualization research. This ability of proof is deeply related to a similar sentiment about photos typically associated with the phrase ``pics or it didn't happen'' \cite{eisele2026pics, lau2023pics}. When explicitly asked whether sharing route outlines provided more of a sense that they had actually done a run than if they had posted text stating they had done a run, P05 responded ``\textit{Yeah, 100\%.}'' P07 added on to this comparison, describing how a route outline additionally increases engagement more than text would: 
\begin{quoting}
    ``\textit{I think the route outline is [...] eye-catching. I'd probably, like, engage with it a little bit more, whereas I might not read [...] all the text.}''
\end{quoting}
Wang et al. also discuss proof in the context of SnapPI stickers, some of which are visualizations \cite{wang2022snappi}. However, they do not discuss the existence of the visualization itself as providing the proof that an activity was real, focusing instead on how \textit{data} provides proof and how that proof was based on insights gleaned from rather than the simple existence of the stickers.}

\subsection{\change{Individuality and Conformity}}

\change{Another novel characteristic to the FDS context is the tension between two influences on posting practices: the desires to express individuality and conform to cultural norms.}

\change{Our participants described the importance of expressing their individuality through how often and what they post. They differed in what activities they found relevant to post. For some, posting about gym workouts was uninteresting, as they used these sessions to support their primary sport; for others, working out in the gym was their primary focus and so they would post them more regularly. If an individual is especially excited about an activity, for example if they've started playing a new sport, they may also be more likely to post about that activity than if it were less novel. The novelty and excitement around posting in general is also a factor which influences post frequency, with people noting that they posted less frequently as they continued to use a platform and gather followers:
\begin{quoting}
    ``\textit{I find that, like, the more followers you have, the less you want to share.}'' (P13)
\end{quoting}
However, participants noted that they would be more interested in posting other sports if they had better visualization support.}

\change{Choices about what information to include or emphasize in a post and how to convey that information also differed based on participant expression. Our participants described how their personal values impact what information they want to share, with the metric of calories burned during an activity being particularly contentious as some individuals saw it as a general descriptor of their effort level while others associated it with disordered eating. The activities that an individual commonly does also impacts their choices, for example with individuals who do not run hilly or mountainous routes finding the elevation gain metric uninteresting. We noted that participants liked to express this information in different ways, often using text, visual components like emojis, and music. P13 mentioned that:
\begin{quoting}
    ``\textit{[they will] take a screenshot of, like, what [they were] listening to, not [...] the playlist, but just [...] a song,}''
\end{quoting}
describing how they use a single song to express how they felt on their activity through their interests and personality. Participants also mentioned expressing themselves and their feelings by intentionally making their posts humourous:
\begin{quoting}
    ``\textit{It's kind of a funny picture, so… I hope they would, you know, get a little laugh}'' (P03).
\end{quoting}}

\remove{\subsection{Conforming to Cultural Norms}}

\change{On the other hand, participants felt pressure to conform to the cultural norms of the platform they were posting on. They also described these norms as an etiquette and expected others on the platform to follow them as well, expressing frustration when they did not. Once again, cultural norms influences how often and what participants posted. In terms of frequency, participants were concerned about posting too frequently, or about posting mundane activities. For example, daily cycling commutes and short walks were viewed negatively and perceived as ``clogging up'' the feed. Meanwhile, some participants felt that it was required to post about activities like social group workouts due to the pressure from the other individuals who participated:
\begin{quoting}
    ``\textit{The people at the Run Club know that I'm running, so they... I feel like it's expected for me to post}'' (P18).
\end{quoting}}

\change{When discussing the ways in which cultural norms influenced their post content, we noted that fitness data sharing did not have one cohesive culture, but was instead divided into subcultures. For example, one participant described how they posted less information over time because they saw others putting in less effort, and they were embarrassed about their effort level. In contrast, another participant described how they began training with a running group at a running track, and noticed how much detail the other runners included in their posts, and then increased the amount of information they included to conform, saying: 
\begin{quoting}
    ``\textit{I noticed that other people in this group also do that, so that's definitely, like, I'm trying to imitate that}" (P12).
\end{quoting} Not all participants chose to conform despite awareness of the social pressure, with one participant noting that they were known as the type to post differently than others.}

\change{The importance of expressing individuality while conforming to cultural norms is a tension that all members of an FDS culture or subculture navigate. While this tension is a novel contribution to visualization research, we believe that it may not be novel to FDS contexts. The desire to be seen as an individual while avoiding clashes with social pressure is certainly not unique to FDS contexts \cite{torelli2006individuality}, and is likely influencing visualization design behaviour in other visualization contexts.}

\change{\subsection{Effort and Desire to Share}}

\change{Participants frequently described their unwillingness to put effort into sharing despite citing many different motivations for them to share and many different goals with what they share. While effort levels differ amongst different individuals, many participants told us that they want sharing to be as low effort as possible and that they would simply not share something if they perceived the process as too much effort. On the surface, this decision not to post when it takes too much effort is surprising given many platforms have automatic upload and sharing features. Digging deeper, we found that our participants perceived different levels of quality and polish as required based on the platform and activity. For platforms, participants described Strava as the least effort to post on, with Instagram's ephemeral Stories filtered to only close friends being more effort, followed by Instagram Stories posted to all followers, then Instagram's perennial Posts. These comparisons were often made explicitly:
\begin{quoting}
    ``\textit{It's a lot more effort [...] for a post, so I'm like, it's easier just to update story-wise}'' (P01).
\end{quoting}
Participants supported this hierarchy with the observations that Instagram is, in general, more ``\textit{performative}'' (P06) than Strava, that Strava posts typically have fewer photos than Instagram posts, that sharing only with those they are close with reduces their fear of being judged, and that they had less desire to share when they have more followers.}

\change{The activity that a person is sharing about also impacts how much effort they feel the post warrants or requires. For example, an individual may decide to put more effort into a post and ``upgrade'' it to a higher-effort platform if they want to share the photos they took:
\begin{quoting}
    ``\textit{If like, it's like a beautiful day, [...] I did a run, and, like, there was… so many beautiful pictures that I took, so that I also took a snapshot of my Strava that posts the… length of the run that I did}'' (P02).
\end{quoting}
Participants also mentioned being more likely to put additional effort into more substantial activities,
\begin{quoting}
    ``\textit{I usually post if I've done a big hike. I think that usually makes it to the post}'' (P01),
\end{quoting}
and more formal or effortful activities,
\begin{quoting}
    ``\textit{I think when I do something big, like, for example, like, if there's a race, that I post on Instagram}'' (P03).
\end{quoting}
These activities contrast more routine or uneventful activities, like solo runs, for which one participants says they ``\textit{don't} [...] edit the post at all'' (P03).}

\change{Some stated a lack of care about others' perceptions, including P16, who said they ``\textit{honestly don't care what anybody thinks about [their] workout}.'' However, it is clear from our study results that, for most, effort is an important consideration when deciding whether and what to share in an FDS context. Since the reasoning for this perceived mandatory effort is based off of influences of social pressure to achieve adequate post quality, we suggest that similar decision trees may exist in other sharing contexts as well.}

\section{Discussion}

We now discuss \change{implications for design,} the potential for our findings to apply to other similar contexts, methodological reflections, and limitations and future work.

\subsection{Implications for Design}

From our \change{interview }study, we draw three implications for the design of \change{FDVS features on FDS} platforms. \change{We draw these design implications from the Desires high-level code, which represents explicitly requested features and is detailed in Table~\ref{tab:properties}, and from reflection on our thematic analysis results.} \change{While these implications are not empirically validated and may not be of interest to all users, they were consistently explicitly and implicitly desired by our participants.}

\subsubsection{More Visualization and Metric Options}

\change{Our participants explicitly expressed the need for more} metric and visualization options \change{in general, and especially} for non-route sports, \change{stating
\begin{quoting}
    ``\textit{More ways to visualize data would be cool}'' (P18)
\end{quoting}
and
\begin{quoting}
    ``\textit{Interviewer: would you be more inclined to post those things if they had more interesting things to show? Like, for example, if you did weightlifting and it had, like, a thing like your swimming goggles, where it shows, like, sets and different types of activities you do. \\
    Participant: Oh, yeah, I totally would.}" (P04).
\end{quoting}
}
The high-level \change{map and heatmap visualizations for activities with location data drive higher engagement with these activities' posts}. {These location-based activities also typically have more and more applicable metrics and visualizations}, including summative numerical descriptions like distance, abstract visual glyphs in the form of route outlines with no geographic map in the background, and concrete visualizations in the form of routes embedded into geographic maps with additional data such as pace embedded in the route line. Sports without routes typically have no visualization options at all, and the included metrics are generic to any activity, such as heart rate or calories burned. To improve the sharing experience of these activities, platforms should consider \remove{activity-specific} metrics, abstract visual glyphs, and concrete visualizations \change{which are activity-specific} to align with the functionality available for route-specific activities. 

\remove{These platforms should also consider additional visualizations and metrics to support comparisons with overall fitness and previous activities. For example, enabling a user to share how their performance compares against recent activities or include metrics that are shown relative to other activities rather than in absolute units would support the types of information sharing that fitness data sharers are already doing. Fitness data platforms do support comparisons to overall fitness and previous activities as part of their personal analysis features, but these analyses are not yet easy to share.}

Our participants also described the usefulness of concreteness. For those who are familiar with a sport, showing information in units such as meters or kilometers is most precise and therefore preferred. However, those who are unfamiliar may better understand if this information were represented in terms of ideas such as flights of stairs or football fields \cite{chevalier2013using}.

\subsubsection{Greater User Control for Metrics}

Our participants described feelings of frustration when discussing the metrics that FDS platforms include in posts. The default metrics are not always the ones that people want to show:
\change{
\begin{quoting}
    ``\textit{[Strava], like, gave me the power level for my [boxing] class. I'm like, I don't care. [...] Like, I feel like it's… I'm not a cyclist, and like… for a boxing class, like, I don't think it was, like, a relevant thing}'' (P02)
\end{quoting}
}
However, the only way to control which metrics are shown, if it is possible at all to change them, is by hiding metrics and then checking what the system replaced them with. These platforms should allow sharers to explicitly choose which metrics they want to share. Sharers should also have control over the granularity of the shared metrics, as a person may be comfortable sharing an overall average but not the minute-by-minute details. Finally, while these platforms provide some control over how metrics are viewed by allowing a user to choose metric or imperial units, they could also enable features such as global hiding of metrics which a user might never want to see, in their posts or others. These increases to user control over metrics would help to alleviate privacy concerns and align posts with the activity, and user values \cite{simpson2017calorie} and goals.

\subsubsection{Increased Ease of Use}

FDS platforms have become increasingly easy to use, as noted by some of our participants, and it is important to maintain this trajectory\change{, as participants ``\textit{do wish that they were a little bit easier to customize}'' (P02).} As platforms implement \change{future features}, they must maintain this ease of use. Some of our participants desired more options for stylistic customization through layout, colour, and font options, but they wanted a small set of options to choose from rather than full flexibility:
\change{
\begin{quoting}
    ``\textit{I feel like curating something from scratch would be a bit too much}'' (P01).
\end{quoting}
}
Sharers may be happy to spend significant amounts of time customizing their posts, for example by changing styles or visible metrics, but the platform should allow users to create templates and profiles which they can design once and then persist for later re-use, as sharers are currently dissuaded by increasing their customization efforts due to frustration about needing to repeat their edits for every activity. None of our participants wanted full design toolkits embedded in the FDS platforms, possibly due to the existing availability of feature-full third-party tools.

\subsection{Extending to Similar Contexts}

There are many areas with established online communities where users share information regularly, one example being video games \cite{rossi2008mmorpg, zhu2021psychology, scully2016re, winger2021super}. Many video games have communities, with huge diversity between the purpose and type of interaction within these communities. Nintendo's \textit{Animal Crossing: New Horizons} enabled people to be social in creative ways, showcasing their individuality through digital creations \cite{zhu2021psychology}. Even in competitive fighting games and speedrunning, players can showcase individuality by finding success with unpopular strategies, with players becoming famous for doing so \cite{hungrybox}. In massive multiplayer online games (MMOs) and team based games, players often hold specific roles within their individual communities and teams, often aligning with the player's personal goals with the game \cite{mondal2022does}. Video game communities also contain cultural norms and pressure to conform, for example with players conforming to popular strategies, learning from others, and changing roles due to social pressure \cite{vorderer2003explaining, rossi2008mmorpg}.

These communities involve a significant amount of information and data sharing about designs, strategies, playthroughs, and even the games themselves, with social comparison being a motivator for sustained engagement. While this context currently involves frequent social sharing of information, many games already visualize information for the users \cite{zammitto2008visualization}, and there are online platforms for sharing data such as speedrunning times in consistent formats. Many of these communities would likely use visualization to communicate specific insights like how fast they were able to accomplish a task. However, in games where the primary challenge is completing a task at all, visualizations may serve as proof of completion, similar to our FDS interview participants. As more visualization tools arise, we may see video game players beginning to curate visualizations of their own game data on a large scale, similar to the FDS context. 

We foresee video game communities, along with other social communities where information sharing is common, becoming more similar to the FDS context over time. If this prediction proves to be true, then they may share traits with the FDS context such as our novel characteristics.

\subsection{Reflections on Methodology}

\change{While we used reflexive thematic analysis \cite{braun2019reflecting}, we were additionally guided by} constructivist grounded theory \change{\cite{charmaz2006constructing}}. \remove{One notable deviation was the authors' domain expertise and preconceptions and knowledge of existing work prior to beginning data collection.} The original form of grounded theory guides users of the methodology to avoid the use of existing knowledge in the form of extant theory and personal domain expertise before data collection and analysis \cite{glaser1998grounded}. However, both Charmaz \cite{charmaz2006constructing} and Furniss et al.~\cite{furniss2011confessions} note that careful use of existing knowledge can support the research without introducing significant bias. As mentioned in Section~\ref{sec:methodology}, it is our proximity to the domain which enabled this project, both in project identification and participant recruitment. We found bracketing to be helpful in reducing the issues with this proximity, and we suggest that others in visualization consider bracketing techniques reflexive journaling and peer debriefing when working on domains in which they have expertise \cite{tufford2012bracketing}.

When conducting studies about design, participants without design expertise may struggle to actively think about what they do and why they do it. Furniss et al.~suggest that researchers overcome this challenge by proposing ideas and and discussing them with participants, rather than following standard grounded theory advice and keeping data collection fully open ended to avoid bias \cite{furniss2011confessions}. This finding aligns with visualization as well, as many people lack expertise in visualization design, and as such struggle to think of potential designs aside from those they are already familiar with. For our interview study, we attempt to combine both guidelines by first asking open-ended questions to gather unbiased participant ideas, get to know the participant, and gain rapport with them, before later providing specific prompts and discussing. We found that this combination enabled us to collect diverse opinions \change{while still prompting participants about} ideas that \change{they} had not considered. 

Finally, we used a modified form of pair interviewing \cite{akbaba2023two}. We switched which interviewer held the driver and navigator roles between each interview as the two interviewers had subtly different domain expertise and providing both interviewers the opportunity to ask different types of questions enabled broader discussions. We also decided to let the driver keep track of the time during the interview, both because they were the one focused on the interview script and to avoid awkward interruptions from the navigator.

\subsection{Limitations and Future Work}

Our study is intentionally grounded in \change{the specific community of Vancouver-based fitness data sharers}, which provides many benefits to the study such as increased rapport with participants and increased understanding of the data through analyst familiarity. We recruited most of our participants through Instagram, which we found was the most popular general social media platform for sharing fitness data in this community based on author familiarity. \change{While our participants did post on fitness-specific social media platforms, } none of our participants mentioned sharing fitness data posts to \remove{other} general social media platforms \change{ other than Instagram}. We also focused on Strava when considering in-app posts, which is the most popular fitness-focused sharing platform. However, it is possible that there are fitness data sharers who use other general and fitness-focused platforms, and that their use differs from that of our participant pool. Further, we did not aim for globally representative results, although the results of our qualitative exploration could enable broader quantitative studies in the future.

\change{Another limitation of our work is that our interview methods relied on self-reported participant recollections, rather than direct observation of actual posting behaviour by researchers, or self-reporting by participants of their thoughts and feelings in the moment of sharing. This method enabled us to be exploratory and to avoid requiring significant interview preparation from the participants and researchers, but may be biased as memories and beliefs can shift in the time between posting and recollection.}

We made observations during our interviews that we determined were outside of the scope of this study, however they may make for interesting future work. We noted that multiple of our participants recorded their fitness data on multiple devices simultaneously, and we were unable to find out why. We also confirmed findings of existing work showing that some people share posts publicly on platforms like Strava but claim to use the platform exclusively for self tracking, which we did not explore further. \change{While our interviews covered both sharing and viewing, as we often needed to discuss viewing to elicit ideas about sharing,} \change{we} focused our analysis on the sharer perspective\change{. However, the viewer perspective may also be interesting to investigate,} for example by showing people others' posts and discussing with them. Finally, we note that our participants did not align in many different ways, and future work could attempt to distill specific traits of fitness data sharers, then conduct a quantitative study to look for correlations between these traits and construct sharer types.

\change{We found that many participants were dissatisfied with the lack of shareable visualization options for activities without location data, especially given the importance of the visualization as proof that the a person did an activity. Our design implication to add other options is based on this observation. However, this visualization gap also led to limitations in our study: our second novel characteristic, about individuality and conformity, and our third novel characteristic, about effort and desire to share, both lack specificity to visualization, and instead discuss fitness data posts more generally. Future work could develop additional visualization options and gather more information on how these characteristics apply to visualizations in fitness data posts.}

\section{Conclusion}

We conducted and thematically analyzed 18 semi-structured interviews to produce a \change{set of novel characteristics arising from our study of the FDS context}. We find that \change{the existence of} visualization has a novel role in this context as a visual stamp of proof that a sharer actually did an activity. We also make suggestions on how to support this role for posts about non-route activities, \change{which currently lack support}. 

Our supplemental materials include our full codebook \remove{containing the codes }with definitions and sample quotes\remove{ and properties with definitions}. They also include a document with additional post and export figures, platform feature collection details and descriptions, saturation details, an explanation of the relationships between \change{different levels of codes}, and a \change{contribution statement} \remove{using the CRediT system }\cite{brand2015beyond}. They also include our interview protocol and script, participant demographic details, and all 18 de-identified interview transcripts. Supplemental materials are available at \url{https://osf.io/rma7w/}.

\section*{Acknowledgments}

The development of this paper took place at the UBC Point Grey campus in the City of Vancouver, which sit on the traditional, ancestral, unceded territory of the Musqueam, Squamish, and Tsleil-Waututh First Nations. This work was supported in part by NSERC DG RGPIN-2024-06401. This study was approved by the UBC BREB (H25-01410)\change{ and informed consent was obtained from all participants}. We thank Steve Kasica, Ryan Smith, Matt Oddo, and Ricky Curry for their feedback on this paper.



\bibliographystyle{IEEEtran}
\bibliography{main.bib}

@string{TVCG = "IEEE Trans. Visualization \& Computer Graphics (TVCG)"}

@STRING{AVI = {Conf. Advanced Visual Interfaces (AVI)}}

@STRING{CHI = {ACM Conf. Human Factors in Computing Systems (CHI)}}

@STRING{DIS = {ACM Conf. Designing Interactive Systems (DIS)}}

@string{BELIV = {IEEE VIS Wkshp. Evaluation and Beyond - Methodological Approaches to Visualization (BELIV)}}

@string{PACM-CHI = {ACM Human-Computer Interaction (PACM-CHI)}}

@string{UBICOMP = {ACM Conf. Pervasive \& Ubiquitous Computing (UbiComp)}}

@string{CSCW = {ACM Conf. Computer Supported Cooperative Work \& Social Computing (CSCW)}}

@string{VIS = {IEEE Conf. Visualization and Visual Analytics (VIS)}}

@string{SOUPS = {Symp. Usable Privacy and Security (SOUPS)}}

@string{TOCHI = {ACM Trans. on Computer-Human Interaction (TOCHI)}}

@string{VISCOMM = {IEEE VIS Wkshp. Visualization for Communication (VisComm)}}

@string{DIS = {Conf. Designing Interactive Systems (DIS)}}

@inproceedings{akbaba2023two,
  title={“{Two} Heads are Better than One”: {P}air-Interviews for Visualization},
  author={Akbaba, Derya and Meyer, Miriah},
  booktitle=VIS,
  pages={206--210},
  year={2023},
  organization={IEEE}
}

@inproceedings{alqhatani2019there,
  title={“{There} is nothing that {I} need to keep secret”: Sharing Practices and Concerns of Wearable Fitness Data},
  author={Alqhatani, Abdulmajeed and Lipford, Heather Richter},
  booktitle=SOUPS,
  pages={421--434},
  year={2019}
}

@inproceedings{bartram2011smart,
  title={Smart homes or smart occupants? {S}upporting aware living in the home},
  author={Bartram, Lyn and Rodgers, Johnny and Woodbury, Rob},
  booktitle={IFIP Conf. Human-Computer Interaction},
  pages={52--64},
  year={2011},
  organization={Springer}
}

@article{baur2010streams,
  title={The streams of our lives: {V}isualizing listening histories in context},
  author={Baur, Dominikus and Seiffert, Frederik and Sedlmair, Michael and Boring, Sebastian},
  journal=TVCG,
  volume={16},
  number={6},
  pages={1119--1128},
  year={2010},
  publisher={IEEE}
}

@article{bhargava2020opportunities,
  title={The opportunities, challenges and obligations of {F}itness {D}ata {A}nalytics},
  author={Bhargava, Yesoda and Nabi, Javaid},
  journal={Procedia Computer Science},
  volume={167},
  pages={1354--1362},
  year={2020},
  publisher={Elsevier}
}

@inproceedings{bowyer2022human,
  title={Human-GDPR interaction: practical experiences of accessing personal data},
  author={Bowyer, Alex and Holt, Jack and Go Jefferies, Josephine and Wilson, Rob and Kirk, David and David Smeddinck, Jan},
  booktitle={Proceedings of the 2022 CHI conference on human factors in computing systems},
  pages={1--19},
  year={2022}
}

@article{brand2015beyond,
  title={Beyond authorship: Attribution, contribution, collaboration, and credit.},
  author={Brand, Amy and Allen, Liz and Altman, Micah and Hlava, Marjorie and Scott, Jo},
  journal={Learned Publishing},
  volume={28},
  number={2},
  year={2015}
}

@article{braun2019reflecting,
  title={Reflecting on reflexive thematic analysis},
  author={Braun, Virginia and Clarke, Victoria},
  journal={Qualitative research in sport, exercise and health},
  volume={11},
  number={4},
  pages={589--597},
  year={2019},
  publisher={Taylor \& Francis}
}

@article{brehmer2013multi,
  title={A multi-level typology of abstract visualization tasks},
  author={Brehmer, Matthew and Munzner, Tamara},
  journal=TVCG,
  volume={19},
  number={12},
  pages={2376--2385},
  year={2013},
  publisher={IEEE}
}

@book{charmaz2006constructing,
  title={Constructing grounded theory: {A} practical guide through qualitative analysis},
  author={Charmaz, Kathy},
  year={2006},
  publisher={Sage}
}

@inproceedings{chetty2011my,
  title={Why is my {I}nternet slow? {M}aking network speeds visible},
  author={Chetty, Marshini and Haslem, David and Baird, Andrew and Ofoha, Ugochi and Sumner, Bethany and Grinter, Rebecca},
  booktitle=CHI,
  pages={1889--1898},
  year={2011}
}

@article{chevalier2013using,
  title={Using concrete scales: {A} practical framework for effective visual depiction of complex measures},
  author={Chevalier, Fanny and Vuillemot, Romain and Gali, Guia},
  journal=TVCG,
  volume={19},
  number={12},
  pages={2426--2435},
  year={2013},
  publisher={IEEE}
}

@inproceedings{consolvo2008activity,
  title={Activity Sensing in the wild: {A} Field Trial of {U}bi{F}it Garden},
  author={Consolvo, Sunny and McDonald, David W and Toscos, Tammy and Chen, Mike Y and Froehlich, Jon and Harrison, Beverly and Klasnja, Predrag and LaMarca, Anthony and LeGrand, Louis and Libby, Ryan and others},
  booktitle=CHI,
  pages={1797--1806},
  year={2008}
}

@inproceedings{costanza2012understanding,
  title={Understanding domestic energy consumption through interactive visualisation: a field study},
  author={Costanza, Enrico and Ramchurn, Sarvapali D and Jennings, Nicholas R},
  booktitle=UBICOMP,
  pages={216--225},
  year={2012}
}

@article{couture2021reflections,
  title={Reflections from the ‘Strava-sphere’: Kudos, community, and (self-) surveillance on a social network for athletes},
  author={Couture, Jesse},
  journal={Qualitative Research in Sport, Exercise and Health},
  volume={13},
  number={1},
  pages={184--200},
  year={2021},
  publisher={Taylor \& Francis}
}

@article{cowley2021invisible,
  title={“Invisible sportswomen”: the sex data gap in sport and exercise science research},
  author={Cowley, Emma S and Olenick, Alyssa A and McNulty, Kelly L and Ross, Emma Z},
  journal={Women in Sport and Physical Activity Journal},
  volume={29},
  number={2},
  pages={146--151},
  year={2021},
  publisher={Human Kinetics}
}

@article{davidson2026spotify,
  title={Spotify {W}arped: Reshaping Personal Informatics via Music Listening, Casual Users, Passive Data and Episodic Reflection},
  author={Davidson, Thomas James and Lee, Ethan and Wall, Emily},
  year={2026},
  journal=CHI,
}

@inproceedings{dias2012interactive,
  title={Interactive exploration of music listening histories},
  author={Dias, Ricardo and Fonseca, Manuel J and Gon{\c{c}}alves, Daniel},
  booktitle=AVI,
  pages={415--422},
  year={2012}
}

@article{eisele2026pics,
  title={Pics or it didn’t happen! EU institutions’ visual communication and user engagement on Facebook and Instagram},
  author={Eisele, Olga and Heidenreich, Tobias and Maares, Phoebe},
  journal={European Journal of Political Research},
  pages={1--26},
  year={2026},
  publisher={Cambridge University Press}
}

@inproceedings{elmqvist2012patterns,
  title={Patterns for visualization evaluation},
  author={Elmqvist, Niklas and Yi, Ji Soo},
  booktitle=BELIV,
  pages={1--8},
  year={2012}
}

@inproceedings{epstein2013fine,
  title={Fine-grained sharing of sensed physical activity: {A} value sensitive approach},
  author={Epstein, Daniel A and Borning, Alan and Fogarty, James},
  booktitle=UBICOMP,
  pages={489--498},
  year={2013}
}

@inproceedings{epstein2015lived,
  title={A lived informatics model of personal informatics},
  author={Epstein, Daniel A and Ping, An and Fogarty, James and Munson, Sean A},
  booktitle=UBICOMP,
  pages={731--742},
  year={2015}
}

@inproceedings{epstein2015nobody,
  title={From "nobody cares" to "way to go!" {A} Design Framework for Social Sharing in Personal Informatics},
  author={Epstein, Daniel A and Jacobson, Bradley H and Bales, Elizabeth and McDonald, David W and Munson, Sean A},
  booktitle=CSCW,
  pages={1622--1636},
  year={2015}
}

@inproceedings{epstein2020yarn,
  title={Yarn: Adding meaning to shared personal data through structured storytelling},
  author={Epstein, Daniel A and Dontcheva, Mira and Fogarty, James and Munson, Sean A},
  booktitle={Graphics Interface 2020},
  year={2020}
}

@phdthesis{evgenieva2024strava,
  title={Strava -- social media or health app? {E}mpirical Study Report.},
  author={Evgenieva, Anna-Maria},
  year={2024},
  school={Strasbourg University}
}

@inproceedings{fan2012spark,
  title={A spark of activity: exploring informative art as visualization for physical activity},
  author={Fan, Chloe and Forlizzi, Jodi and Dey, Anind K},
  booktitle=UBICOMP,
  pages={81--84},
  year={2012}
}

@article{felczak2025mobile,
  title={Mobile health cycling: {H}ow {E}astern {E}uropean amateur cycling enthusiasts frame their experiences with {Z}wift and {S}trava},
  author={Felczak, Mateusz and Filiciak, Miros{\l}aw},
  journal={Intl. Review for the Sociology of Sport},
  pages={10126902251333566},
  year={2025},
  publisher={SAGE Publications Sage UK: London, England}
}

@article{fiesler2018participant,
  title={“Participant” perceptions of Twitter research ethics},
  author={Fiesler, Casey and Proferes, Nicholas},
  journal={Social Media+ Society},
  volume={4},
  number={1},
  pages={2056305118763366},
  year={2018},
  publisher={SAGE Publications Sage UK: London, England}
}

@article{franken2023kudos,
  title={Kudos make you run! {H}ow runners influence each other on the online social network {S}trava},
  author={Franken, Rob and Bekhuis, Hidde and Tolsma, Jochem},
  journal={Social Networks},
  volume={72},
  pages={151--164},
  year={2023},
  publisher={Elsevier}
}

@inproceedings{furniss2011confessions,
  title={Confessions from a grounded theory {PhD}: experiences and lessons learnt},
  author={Furniss, Dominic and Blandford, Ann and Curzon, Paul},
  booktitle=CHI,
  pages={113--122},
  year={2011}
}

@inproceedings{ge2025avec,
  title={{AVEC}: An Assessment of Visual Encoding Ability in Visualization Construction},
  author={Ge, Lily W and Cui, Yuan and Kay, Matthew},
  booktitle=CHI,
  pages={1--16},
  year={2025}
}

@book{glaser1998grounded,
  title={Discovery of Grounded Theory: {S}trategies for Qualitative Research},
  author={Glaser, Barney G and Strauss, Anselm L},
  year={1999},
  publisher={Routledge},
}

@article{guest2006many,
  title={How many interviews are enough? An experiment with data saturation and variability},
  author={Guest, Greg and Bunce, Arwen and Johnson, Laura},
  journal={Field methods},
  volume={18},
  number={1},
  pages={59--82},
  year={2006},
  publisher={Sage Publications Sage CA: Thousand Oaks, CA}
}

@inproceedings{gui2017fitness,
  title={When fitness meets social networks: {I}nvestigating fitness tracking and social practices on {W}e{R}un},
  author={Gui, Xinning and Chen, Yu and Caldeira, Clara and Xiao, Dan and Chen, Yunan},
  booktitle=CHI,
  pages={1647--1659},
  year={2017}
}

@article{huang2014personal,
  title={Personal visualization and personal visual analytics},
  author={Huang, Dandan and Tory, Melanie and Aseniero, Bon Adriel and Bartram, Lyn and Bateman, Scott and Carpendale, Sheelagh and Tang, Anthony and Woodbury, Robert},
  journal=TVCG,
  volume={21},
  number={3},
  pages={420--433},
  year={2014},
  publisher={IEEE}
}

@article{johnston2019motivating,
  title={Motivating exercise through social media: {I}s a picture always worth a thousand words?},
  author={Johnston, Caitlyn and Davis, William E},
  journal={Psychology of Sport and Exercise},
  volume={41},
  pages={119--126},
  year={2019},
  publisher={Elsevier}
}

@article{lau2023pics,
  title={"Pics or It Didn't Happen" and" Show Me the Receipts": A Folk Evidentiary Rule},
  author={Lau, Timothy},
  journal={Vand. L. Rev.},
  volume={76},
  pages={1681},
  year={2023},
  publisher={HeinOnline}
}

@article{li2012using,
  title={Using context to reveal factors that affect physical activity},
  author={Li, Ian and Dey, Anind K and Forlizzi, Jodi},
  journal = TOCHI,
  volume={19},
  number={1},
  pages={1--21},
  year={2012},
  publisher={ACM New York, NY, USA}
}

@article{li2026platform,
  title={From Platform Data to Personal Insight: {H}ow Users Make Sense of and Reflect on Personalized Social Media Annual Recaps},
  author={Li, Wenqi and Zhang, Jinghan and Ma, Junyang and Zhang, Pengyi},
  year={2026},
  journal=CHI
}

@article{liang2016sleepexplorer,
  title={Sleep{E}xplorer: a visualization tool to make sense of correlations between personal sleep data and contextual factors},
  author={Liang, Zilu and Ploderer, Bernd and Liu, Wanyu and Nagata, Yukiko and Bailey, James and Kulik, Lars and Li, Yuxuan},
  journal={Personal and Ubiquitous Computing},
  volume={20},
  number={6},
  pages={985--1000},
  year={2016},
  publisher={Springer}
}

@inproceedings{lin2024functional,
  title={Functional design requirements to facilitate menstrual health data exploration},
  author={Lin, Georgianna and Lessard, Pierre-William and Le, Minh Ngoc and Li, Brenna and Chevalier, Fanny and Truong, Khai N and Mariakakis, Alex},
  booktitle=CHI,
  pages={1--15},
  year={2024}
}

@article{lowe2018thumbs,
  title={Thumbs up: {A} thematic analysis of image-based posting and liking behaviour on social media},
  author={Lowe-Calverley, Emily and Grieve, Rachel},
  journal={Telematics and Informatics},
  volume={35},
  number={7},
  pages={1900--1913},
  year={2018},
  publisher={Elsevier}
}

@incollection{manning2017vivo,
  title={In vivo coding},
  author={Manning, Jimmie},
  booktitle={The international encyclopedia of communication research methods},
  editors = {J. Matthes and C.S. Davis and R.F. Potter},
  pages={18},
  year={2017}
}

@inproceedings{mondal2022does,
  title={Does A Support Role Player really Create Difference towards Triumph? {A}nalyzing Individual Performances of Specific Role Players to Predict Victory in {L}eague of {L}egends},
  author={Mondal, Joyanta Jyoti and Zahin, Abrar and Manab, Meem Arafat and Hasan, Mohammad Zahidul},
  booktitle={Intl. Conf. Computer \& Information Technology (ICCIT)},
  pages={768--773},
  year={2022},
  organization={IEEE}
}

@article{moore2021exploring,
  title={Exploring the Personal Informatics Analysis Gap: “{T}here's a Lot of Bacon”},
  author={Moore, Jimmy and Goffin, Pascal and Wiese, Jason and Meyer, Miriah},
  journal=TVCG,
  volume={28},
  number={1},
  pages={96--106},
  year={2021},
  publisher={IEEE}
}

@book{munzner2014visualization,
  title={Visualization Analysis and Design},
  author={Munzner, Tamara},
  publisher = {CRC Press},
  year={2014},
}

@article{rivers2020strava,
  title={Strava as a discursive field of practice: {T}echnological affordances and mediated cycling motivations},
  author={Rivers, Damian J},
  journal={Discourse, Context \& Media},
  volume={34},
  pages={100345},
  year={2020},
  publisher={Elsevier}
}

@article{rossi2008mmorpg,
  title={{MMORPG} Guilds as Online Communities-Power, Space and Time: {F}rom Fun to Engagement in Virtual Worlds},
  author={Rossi, Luca},
  journal={Space and Time: From Fun to Engagement in Virtual Worlds (August 28, 2008)},
  publisher = {SSRN},
  year={2008}
}

@article{russell2023if,
  title={“{I}f It's not on {S}trava It Didn’t Happen”: {P}erceived Psychosocial Implications of {S}trava use in Collegiate Club Runners},
  author={Russell, Hayley C and Potts, Charlie and Nelson, Emma},
  journal={Recreational Sports Journal},
  volume={47},
  number={1},
  pages={15--25},
  year={2023},
  publisher={SAGE},
}

@article{sanchez2023effect,
  title={The effect of interest and attitude on public comprehension of news with data visualization},
  author={S{\'a}nchez-Holgado, Patricia and Arcila-Calder{\'o}n, Carlos and Fr{\'\i}as-V{\'a}zquez, Maximiliano},
  journal={Frontiers in Communication},
  volume={8},
  pages={1064184},
  year={2023},
  publisher={Frontiers Media SA}
}

@phdthesis{scully2016re,
  title={Re-curating the accident: {S}peedrunning as community and practice},
  author={Scully-Blaker, Rainforest},
  year={2016},
  school={Concordia University}
}

@article{seidman2013self,
  title={Self-presentation and belonging on {F}acebook: {H}ow personality influences social media use and motivations},
  author={Seidman, Gwendolyn},
  journal={Personality and individual differences},
  volume={54},
  number={3},
  pages={402--407},
  year={2013},
  publisher={Elsevier}
}

@article{simpson2017calorie,
  title={Calorie counting and fitness tracking technology: {A}ssociations with eating disorder symptomatology},
  author={Simpson, Courtney C and Mazzeo, Suzanne E},
  journal={Eating behaviors},
  volume={26},
  pages={89--92},
  year={2017},
  publisher={Elsevier}
}

@mastersthesis{smith2014mobile,
  title={Mobile interactive fitness technologies and the recreational experience of bicycling: {A} phenomenological exploration of the {S}trava community},
  author={Smith, William R},
  year={2014},
  school={Clemson University}
}

@inproceedings{solen2022scoping,
  title={Scoping the future of visualization literacy: {A} review},
  author={Solen, Mara},
  year={2022},
  booktitle = VISCOMM,
}

@article{solen2024delve,
  title={{DeLVE} into {E}arth's Past: {A} Visualization-Based Exhibit Deployed Across Multiple Museum Contexts},
  author={Solen, Mara and Sultana, Nigar and Lukes, Laura and Munzner, Tamara},
  journal=TVCG,
  volume={31},
  number={1},
  pages={952--961},
  year={2024},
  publisher={IEEE}
}

@article{sung2016we,
  title={Why we post selfies: {U}nderstanding motivations for posting pictures of oneself},
  author={Sung, Yongjun and Lee, Jung-Ah and Kim, Eunice and Choi, Sejung Marina},
  journal={Personality and Individual Differences},
  volume={97},
  pages={260--265},
  year={2016},
  publisher={Elsevier}
}

@article{tong2018storytelling,
  title={Storytelling and visualization: {A}n extended survey},
  author={Tong, Chao and Roberts, Richard and Borgo, Rita and Walton, Sean and Laramee, Robert S and Wegba, Kodzo and Lu, Aidong and Wang, Yun and Qu, Huamin and Luo, Qiong and others},
  journal={Information},
  volume={9},
  number={3},
  pages={65},
  year={2018},
  publisher={MDPI}
}

@article{torelli2006individuality,
  title={Individuality or conformity? The effect of independent and interdependent self-concepts on public judgments},
  author={Torelli, Carlos J},
  journal={Journal of Consumer Psychology},
  volume={16},
  number={3},
  pages={240--248},
  year={2006},
  publisher={Elsevier}
}

@article{tufford2012bracketing,
  title={Bracketing in qualitative research},
  author={Tufford, Lea and Newman, Peter},
  journal={Qualitative social work},
  volume={11},
  number={1},
  pages={80--96},
  year={2012},
  publisher={SAGE},
}

@inproceedings{vorderer2003explaining,
  title={Explaining the enjoyment of playing video games: the role of competition},
  author={Vorderer, Peter and Hartmann, Tilo and Klimmt, Christoph},
  booktitle={Intl. Conf. Entertainment Computing},
  pages={1--9},
  year={2003}
}

@article{wang2022snappi,
  title={Snap{PI}: Understanding everyday use of personal informatics data stickers on ephemeral social media},
  author={Wang, Dennis and Chheang, Marawin and Ji, Siyun and Mohta, Ryan and Epstein, Daniel A},
  journal=PACM-CHI,
  volume={6},
  number={CSCW2},
  pages={1--27},
  year={2022},
  publisher={ACM New York, NY, USA}
}

@article{wang2024exploring,
  title={Exploring activity-sharing response differences between broad-purpose and dedicated online social platforms},
  author={Wang, Dennis and Eng, Jocelyn and Turpitka, Mykyta and Epstein, Daniel A},
  journal=PACM-CHI,
  volume={8},
  number={CSCW2},
  pages={1--37},
  year={2024},
  publisher={ACM New York, NY, USA}
}

@article{westlake2020if,
  title={"{I}f you see me collapse, pause my {S}trava": {B}iopower, fitness data, and the anxious online performance of the fit body},
  author={E.J. Westlate},
  journal={Sporting Performances: Politics in Play},
  year={2020},
  publisher={Routledge},
  pages={130--147}
}

@mastersthesis{williams2012king,
  title={King of the {M}ountain: {A} rapid ethnography of {S}trava cycling},
  author={Williams, Alison M},
  year={2012},
  school={University College London}
}

@mastersthesis{winger2021super,
  title={Super Smash Bros. Melee: {P}erformance as Community Survival},
  author={Winger, Nicholas James},
  year={2021},
  school={State University of New York at Buffalo}
}

@inproceedings{wozniak2017soil,
  title={Soil, rock, and snow: {O}n designing for information sharing in outdoor sports},
  author={Wozniak, Pawe{\l} W and Fedosov, Anton and Mencarini, Eleonora and Knaving, Kristina},
  booktitle=DIS,
  pages={611--623},
  year={2017}
}

@inproceedings{zammitto2008visualization,
  title={Visualization techniques in video games},
  author={Zammitto, Veronica},
  booktitle={Electronic Visualisation and the Arts (EVA)},
  year={2008},
  organization={BCS Learning \& Development}
}

@article{zhu2021psychology,
  title={The psychology behind video games during {COVID}-19 pandemic: {A} case study of {A}nimal {C}rossing: {N}ew {H}orizons},
  author={Zhu, Lin},
  journal={Human Behavior and Emerging Technologies},
  volume={3},
  number={1},
  pages={157--159},
  year={2021},
  publisher={Wiley Online Library}
}

@misc{hungrybox,
  title = {Smasher: {Hungrybox}},
  howpublished = {\url{https://www.ssbwiki.com/Smasher:Hungrybox}},
  note = {Accessed: 2026-04-02}
}

\def\interBioSpace{-33pt} 
\vspace{\interBioSpace}
\begin{IEEEbiography}[{\includegraphics[width=1in,height=1.25in,clip,keepaspectratio]{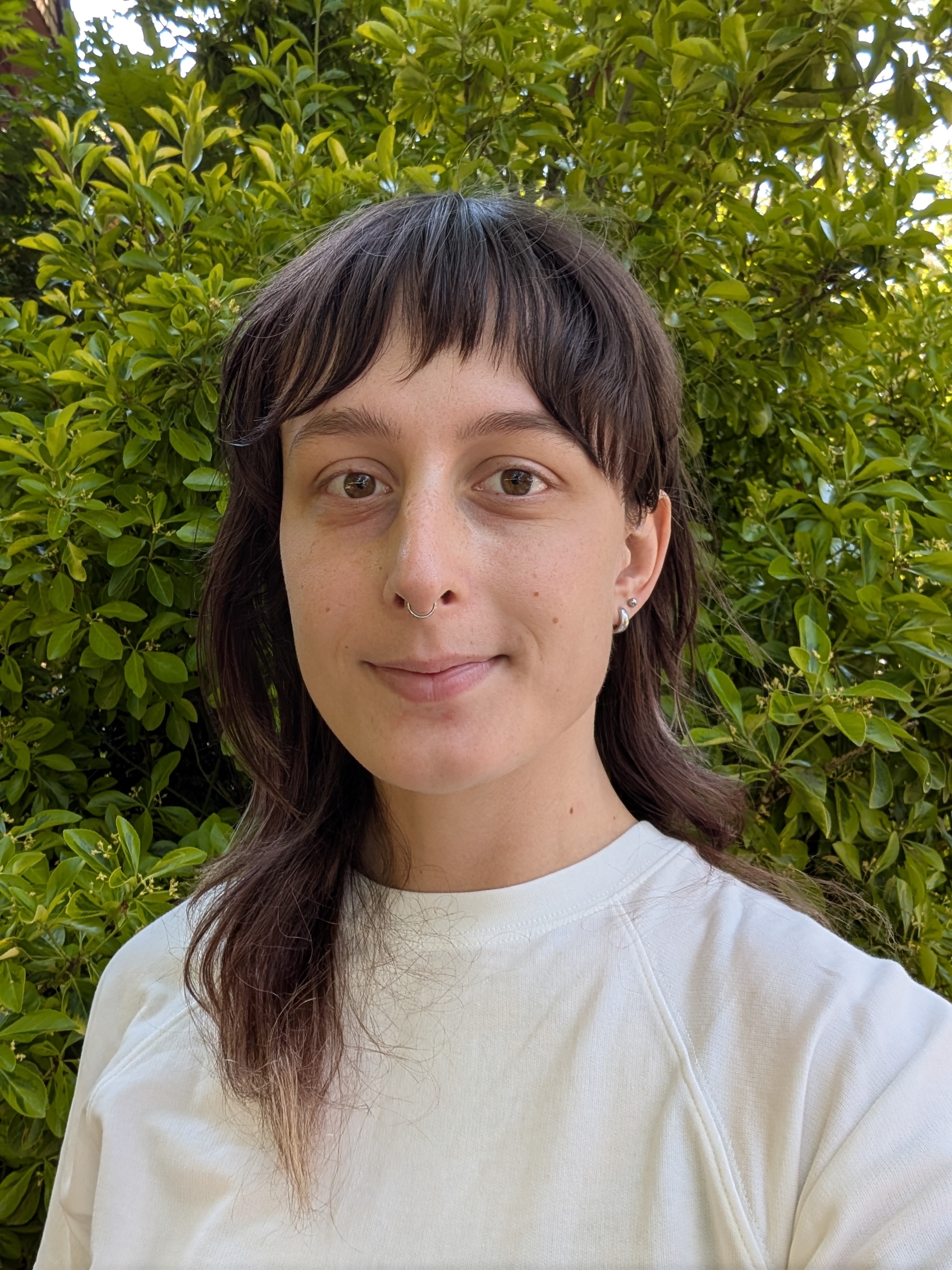}}]{Mara Solen} is a PhD candidate at the University of British Columbia, supervised by Tamara Munzner. Her current research focus is on visualization, focusing on bridging between silos in visualization research, such as between analysis-focused and communication-focused literature, and studying underexplored visualization contexts, such as social uses and reflective uses.
\end{IEEEbiography}

\vspace{\interBioSpace}
\begin{IEEEbiography}[{\includegraphics[width=1in,height=1.25in,clip,keepaspectratio]{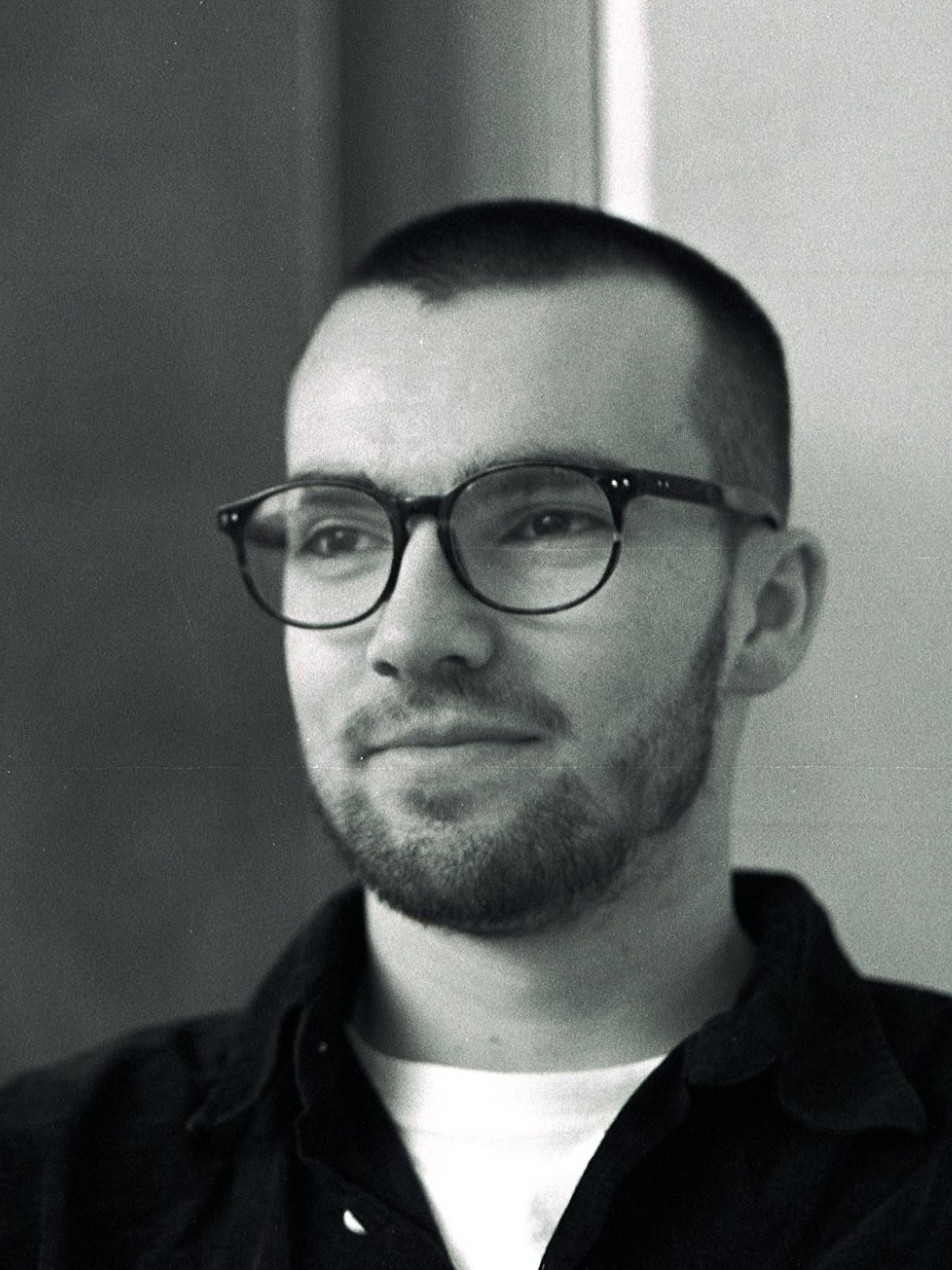}}]{Thomas James Davidson} is a PhD student at the University of Emory, supervised by Emily Wall. His current research centres on broadening the audience of personal informatics with a focus on visualisation related to personal data and how this can increase the agency of more casual users in this space.
\end{IEEEbiography}

\vspace{\interBioSpace}
\begin{IEEEbiography}[{\includegraphics[width=1in,height=1.25in,clip,keepaspectratio]{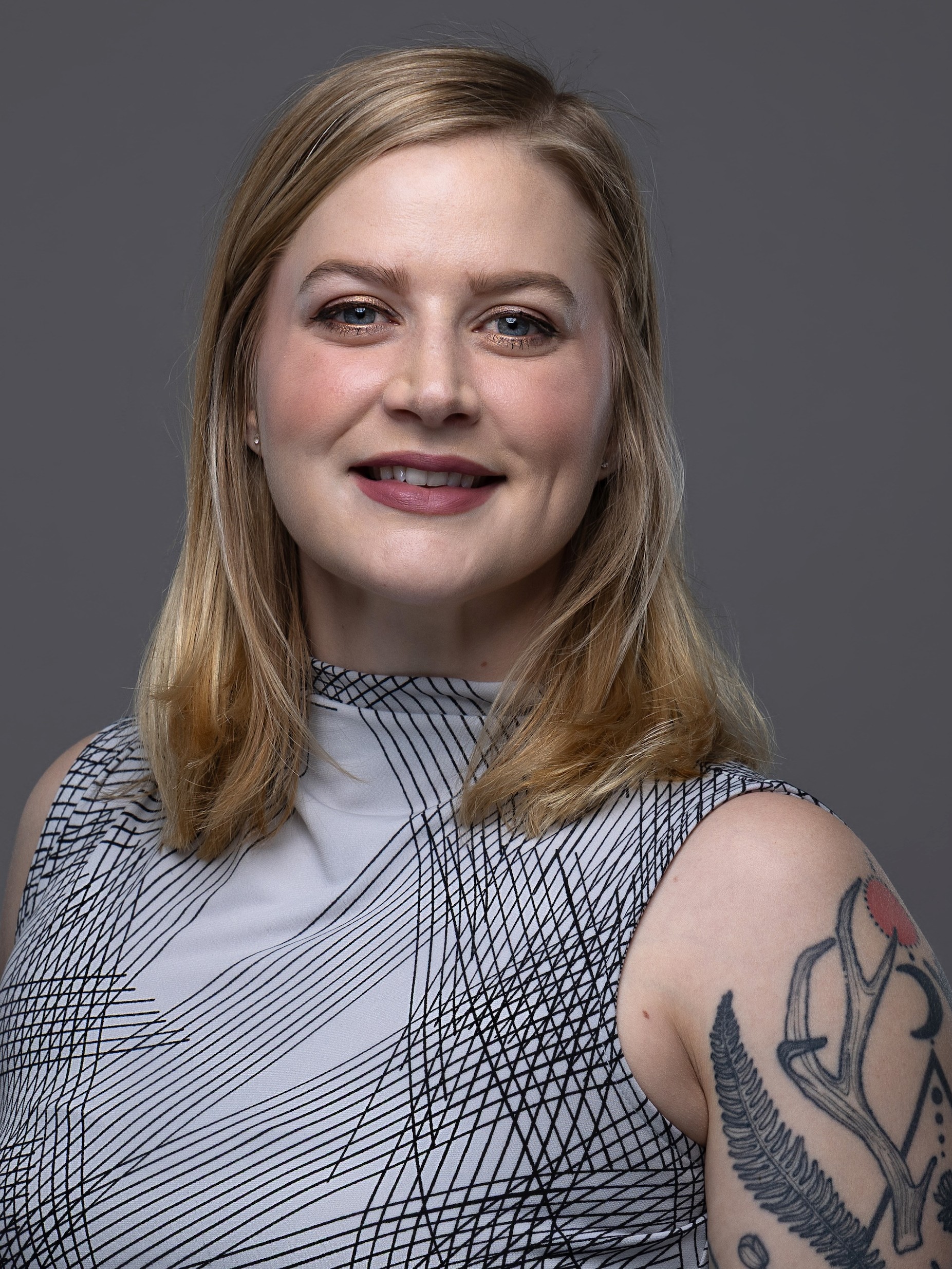}}]{Emily Wall} is an Assistant Professor at Emory University, where she directs the Cognition and Visualization Lab. She completed her Ph.D. in Computer Science from Georgia Tech. Her research interests span decision making and reflection with data and visualization, which has been funded by an NSF CAREER Award on Promoting Metacognition in Visual Analytics. 

\end{IEEEbiography}

\vspace{\interBioSpace}
\begin{IEEEbiography}[{\includegraphics[width=1in,height=1.25in,clip,keepaspectratio]{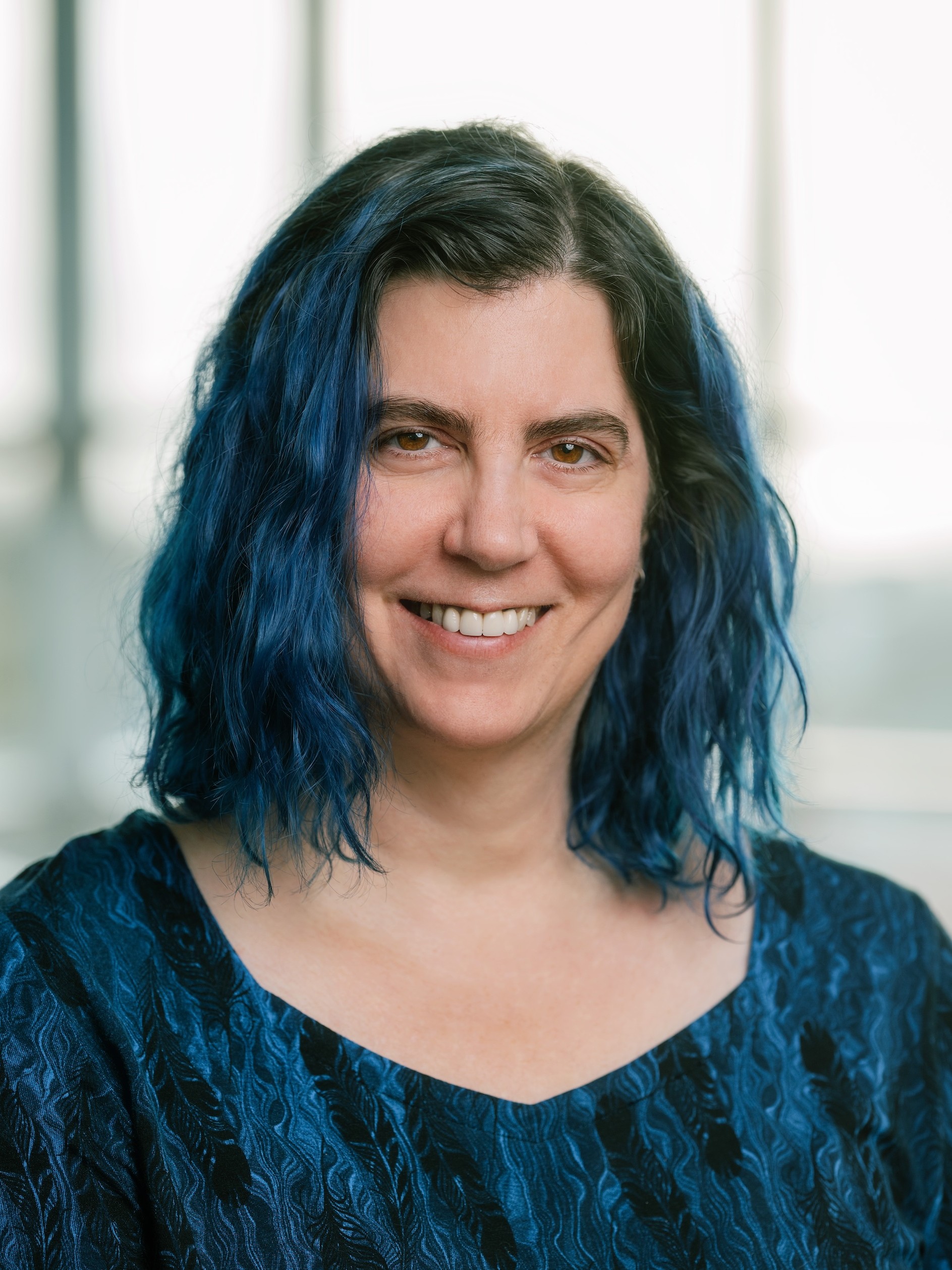}}]{Tamara Munzner} (IEEE Fellow) received the PhD degree from Stanford. She is currently a professor with the University of British Columbia. She has worked on visualization projects in a broad range of application domains from genomics to journalism. Her book Visualization Analysis and Design is heavily used worldwide, and she was the recipient of the IEEE VGTC Visualization Technical Achievement Award.

\end{IEEEbiography}

\vfill

\end{document}